\documentclass[preprint,showpacs,superscriptaddress,preprintnumbers,amsmath,amssymb]{revtex4}
\usepackage{graphicx}
\usepackage{dcolumn}
\usepackage{bm}

\begin{document}


\title{Entanglement swapping secures multiparty quantum
communication}

\author{Juhui Lee}\email{salier97@sookmyung.ac.kr}
\affiliation{
 Department of Physics,
 Sookmyung Women's University,
 Seoul 140-742, Korea
}
\affiliation{
 School of Computational Sciences,
 Korea Institute for Advanced Study,
 Seoul 130-722, Korea
}
\author{Soojoon Lee}\email{level@kias.re.kr}
\affiliation{
 School of Computational Sciences,
 Korea Institute for Advanced Study,
 Seoul 130-722, Korea
}
\author{Jaewan Kim}\email{jaewan@kias.re.kr}
\affiliation{
 School of Computational Sciences,
 Korea Institute for Advanced Study,
 Seoul 130-722, Korea
}
\author{Sung Dahm Oh}\email{sdoh@sookmyung.ac.kr}
\affiliation{
 Department of Physics,
 Sookmyung Women's University,
 Seoul 140-742, Korea
}

\date{\today}

\begin{abstract}
Extending the eavesdropping strategy
devised by Zhang, Li and Guo [Phys. Rev. A \textbf{63}, 036301 (2001)],
we show that the multiparty quantum communication protocol based on entanglement swapping,
which was proposed by Cabello [quant-ph/0009025],
is not secure.
We modify the protocol
so that entanglement swapping can secure multiparty quantum communication,
such as multiparty quantum key distribution and quantum secret sharing of classical information,
and show that the modified protocol is secure against the Zhang-Li-Guo's strategy for eavesdropping
as well as the basic intercept-resend attack.
\end{abstract}

\pacs{03.67.Dd, 03.65.Ta, 03.67.Hk}
\maketitle

\section{Introduction}\label{sec:Introduction}
Entanglement has been considered as one of the most important resources
for quantum information processing and quantum communication
including quantum key distribution (QKD) and quantum secret sharing (QSS).
Entanglements can be obtained from several methods,
one of which is to utilize a projective measurement on multi-particles.
If more than two particles can be handled then
a projective measurement on multi-particles such as a Bell-state measurement
allows the state of the particles to be entangled,
although the particles have never been entangled with one another.

We now consider the following four-particle state,
\begin{eqnarray}
&&(a_1, a_2)_{12}\otimes (a_3, a_4)_{34}\nonumber\\
&&=
\frac{1}{d}\sum_{k,l=0}^{d-1} \omega_d^{kl}
(a_1-k, a_4+l)_{14}\otimes (a_3+k, a_2-l)_{32}.\nonumber \\
\label{eq:entanglement_swapping}
\end{eqnarray}
Here the subscript represents the index of each particle,
each $a_j$ is an integer running from $0$ to $d-1$,
$\omega_{d}=\exp(2\pi i/d)$, and
\begin{equation}
(u,v)_{xy}
=\frac{1}{\sqrt{d}}\sum_{j=0}^{d-1} \omega_d^{ju}
\left|j\right\rangle_x\otimes \left|j+v\right\rangle_y,
\label{eq:two_particle_entanglement}
\end{equation}
which is one of $d$-dimensional Bell states.
The state in Eq.~(\ref{eq:entanglement_swapping}) has
mutually entangled particles 1 and 2, and mutually entangled particles 3 and 4.
There does not exist any entanglement between particles 1 and 4,
nor between particles 2 and 3 in the state.

However, after the measurement on particles 1 and 4 (or 2 and 3)
by the observable with eigenstates $(a_1-k, a_4+l)_{14}$ [or
$(a_3+k, a_2-l)_{32}$] for $0\le k, l \le d-1$, the state in
Eq.~(\ref{eq:entanglement_swapping}) becomes a product state of
the form $(a_1-k, a_4+l)_{14}\otimes (a_3+k, a_2-l)_{32}$, whose
particles 1 and 4, and particles 2 and 3 are mutually entangled,
respectively. This interesting phenomenon is called the {\em
entanglement swapping} (ES), which was originally proposed by
\.{Z}ukowski {\em et al.}~\cite{ZZHE}, was generalized to
multipartite quantum systems by Zeilinger {\em et al.}~\cite{ZHWZ}
and Bose {\em et al.}~\cite{BVK} independently, and was
experimentally realized by Pan {\em et al.}~\cite{PBWZ}.
Furthermore, the generalizations of ES for multipartite and
arbitrary dimensional quantum systems were also presented
\cite{BB,KBB}. These generalizations possess the more significant
meaning in view of the fact that quantum communications in higher
dimensional quantum systems are more secure than in two qubits
\cite{KGZMZ}.
On this account, we consider $d$-dimensional quantum systems,
called qudits, in this paper.

Recently, several practical applications of ES,
such as the purification protocols~\cite{BVK2,HS,SJG}
and the cryptographic protocols~\cite{Cabello,Cabello2,Cabello1,ZLG,KBB},
have been proposed.
In particular,
the ES-based two-party QKD protocol, presented by Cabello~\cite{Cabello,Cabello1,ZLG},
permits the secure key distribution,
whose structure is
different from those of the previous QKD protocols~\cite{BB84,B92,Ekert91}.
In addition,
the ES-based multiparty quantum communication consisting of
multiparty QKD and QSS protocols~\cite{HBB,KKI}
was suggested~\cite{Cabello2,KBB}.
%
However, this multiparty quantum communication 
can be shown to be insecure by means of
a variant of Zhang, Li, and Guo's eavesdropping strategy~\cite{ZLG}.
We call this eavesdropping strategy
the {\em Zhang-Li-Guo-type} (ZLG-type) attack.
In this paper,
we present the ZLG-type attack applied to the multiparty quantum communication protocol,
and show that the protocol is not secure against the attack.
Modifying the protocol
in the method analogous to Cabello's modification for the ES-based two-party QKD~\cite{Cabello2},
we also show that there exists a multiparty quantum communication protocol
which is secure against the ZLG-type attack
as well as the basic intercept-resend attack.

This paper is organized as follows.
In Sec.~\ref{sec:Cabello_protocol}
we review the original Cabello's protocol.
In Sec.~\ref{sec:ZLG}
we show that the protocol is not secure
by employing the ZLG-type attack.
In Sec.~\ref{sec:Modified_protocol}
we present the modified protocol which is secure against
not only the basic intercept-resend attack but also the ZLG-type attack.
Finally, in Sec.~\ref{sec:Conclusions} we summarize our results.


\section{Cabello's protocol for multiparty quantum communication 
}\label{sec:Cabello_protocol}

Cabello~\cite{Cabello1} extended
the ES-based QKD protocol between two parties~\cite{Cabello2}
into multiparty quantum communication.
In this section,
we review the extended protocol for the three-party quantum communication~\cite{Cabello1}.
Since the three-party quantum communication
can easily be generalized to the multiparty case,
we consider only the three-party case in this paper.

In the beginning, Alice prepares one of $d$-dimensional Bell states
and one of $d$-dimensional Greenberger-Horne-Zeilinger (GHZ) states~\cite{GHZ},
and Bob (Carol) prepares one of $d$-dimensional Bell states.
In the protocol,
we assume that the information on the initial state 
is publicly known.
We now remark the following two formulas
for swapping Bell states and GHZ states:
\begin{align}
   (u_1, u_2)_{12} \otimes &(v_1 , v_2, v_3)_{345} \nonumber\\
   = {\frac{1}{d}}\sum_{k,l=0}^{d-1}&\omega_{d}^{k l}(v_1 -k , u_2 +l)_{32}\nonumber\\
   &\otimes(u_1 +k, v_2 -l, v_3 -l)_{145},
  \label{eq:ES2}
\end{align}
and
\begin{align}
   (u_1, u_2)_{12} \otimes &(v_1 , v_2, v_3)_{345} \nonumber\\
   = {\frac{1}{d}}\sum_{k,l=0}^{d-1}&\omega_{d}^{k l}(u_1 -k , v_3 +l)_{15}\nonumber\\
   &\otimes(v_1 +k, v_2 , u_2 -l)_{342},
  \label{eq:ES3}
\end{align}
where $(v_1,v_2,v_3)_{xyz}=\frac{1}{\sqrt{d}}
\sum_{j=0}^{d-1} \omega_{d}^{j v_1}|j, j+v_2, j+v_3\rangle_{xyz}$,
which is a GHZ state.

By exploiting the above two formulas,
the process can be described as follows.
\begin{itemize}
\item[(a)]
Alice 
makes a Bell measurement
on one qudit of the Bell state and one qudit of the GHZ state,
and then
Alice sends one of the other qudits of the GHZ state to
Bob (Carol).

Let
$( a_1, a_2)_{\rm  A}\otimes (g_1, g_2, g_3)_{\rm  A}
\otimes (b_1, b_2)_{\rm  B} \otimes (c_1, c_2)_{\rm  C}$
be the given initial state.
Here, the subscripts represent the indices of the parties manipulating the state.
If we let $(g_1-k, a_2+l)_{\rm A}$ be the result of Alice's measurement
for some $k$ and $l$,
then we obtain from Eq.~(\ref{eq:ES2})
that the initial state becomes
$(g_1-k, a_2+l)_{\rm A}\otimes (a_1+k, g_2-l, g_3-l)_{\rm ABC}
\otimes (b_1, b_2)_{\rm B} \otimes (c_1, c_2)_{\rm  C}$.
\item[(b)] Bob (Carol) performs a Bell measurement
on one qudit of his (her) Bell state and the qudit received from Alice.
After the measurements,
Bob (Carol) sends Alice his (her) qudit which is not measured.

For some $m_i$ and $n_i$, 
we let $(b_1-m_1, g_2-l-m_2)_{\rm  B}$ [$(c_1-n_1, g_3-l-n_2)_{\rm  C}$]
be the measurement result of Bob (Carol).
Then it follows from Eq.~(\ref{eq:ES3})
that the resulting state is
$(g_1-k, a_2 +l)_{\rm A}
\otimes (a_1+k+m_1+n_1, b_2+m_2 , c_2+n_2)_{\rm  A}
\otimes (b_1 -m_1, g_2-l-m_2)_{\rm B}
\otimes (c_1 -n_1, g_3 -l-n_2)_{\rm C}$.
\item[(c)] Alice measures the GHZ state,
$(a_1+k+m_1+n_1, b_2+m_2 , c_2+n_2)_{\rm  A}$,
which consists of her own qudit and the qudits received from Bob and Carol,
and then publicly announces the result of the measurement.
Then the final state becomes
\begin{eqnarray}
&&(g_1 -k, a_2 +l)_{\rm A} \nonumber\\
&&\otimes~(a_1 +k+m_1+n_1, b_2+m_2 , c_2+n_2)_{\rm  Pub}\nonumber\\
&&\otimes~(b_1 -m_1, g_2-l-m_2)_{\rm B} \nonumber\\
&&\otimes~(c_1 -n_1, g_3 -l-n_2)_{\rm C}.
\label{eq:original_final_state}
\end{eqnarray}
The correlation in (\ref{eq:original_final_state})
allows multiparty QKD
and QSS of classical information,
simultaneously.
\end{itemize}
 For the secure communication,
legal users check the eavesdropping
by comparing random subsequences of their key strings with the original correlations.

We note that the basic intercept-resend attacks
analogous to the attack presented in~\cite{Cabello,Cabello1},
may destroy the correlations among the keys,
since any Eve's attempt to acquire the keys may
change the result of the public announcement
in an unpredictable way.
Therefore, Cabello's protocol is secure against
such kind of attacks.


\section{Zhang-Li-Guo-type attack on Cabello's protocol}\label{sec:ZLG}

\begin{figure}[ch] 
\includegraphics[angle=-90,scale=1.00,width=.80\textwidth]{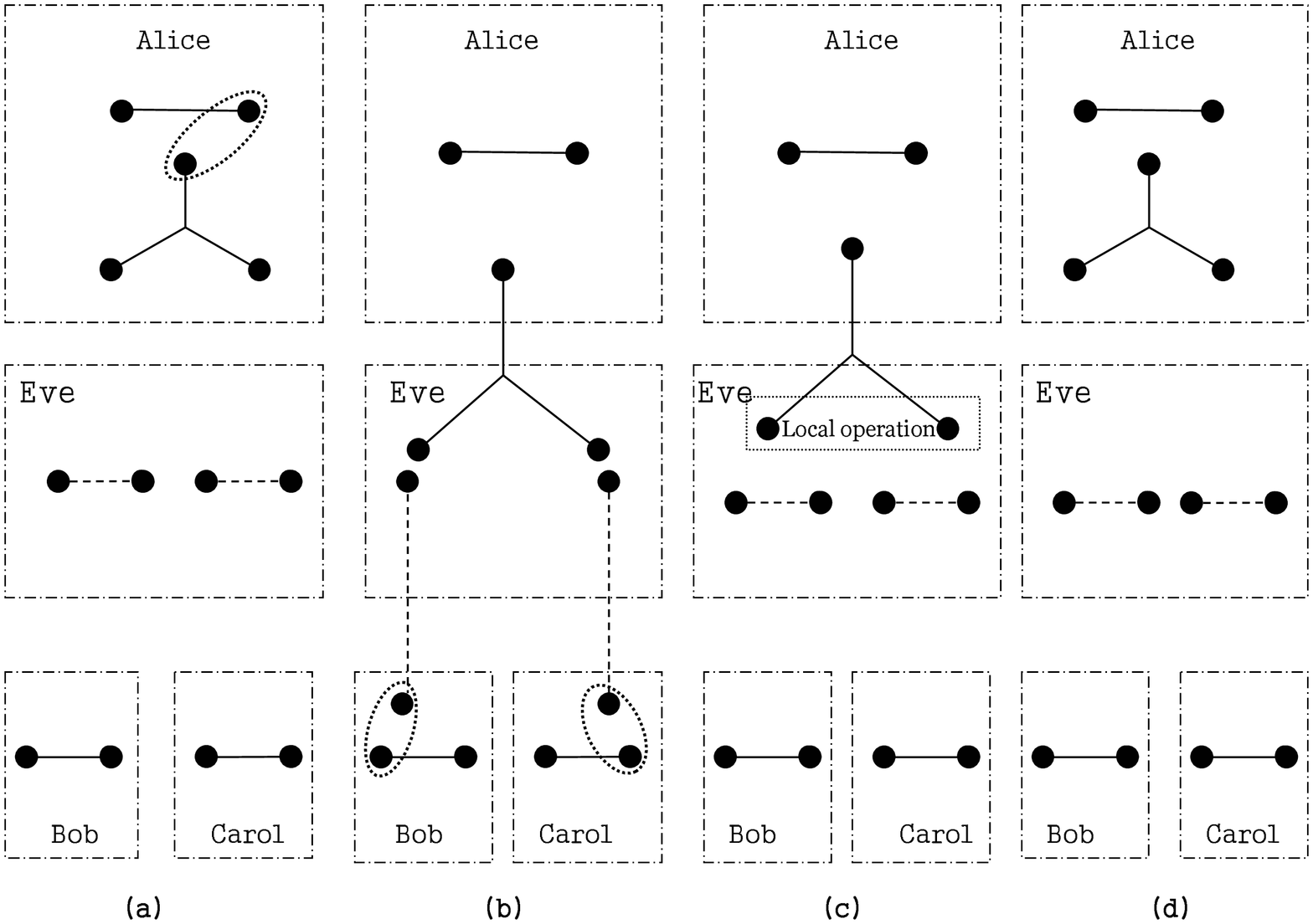}
\caption{\label{fig:fi1}
The ZLG-type attack on Cabello's protocol for the three-party quantum communication:
The solid lines and the dashed lines represent
the entanglement given to legal users and Eve, respectively.
The dotted ellipses and the dotted box represent
performing Bell measurements
and applying the local operation in Eq.~(\ref{eq:local_operation}), respectively.}
\end{figure}

Zhang, Li and Guo~\cite{ZLG} proposed a special strategy for eavesdropping,
called the ZLG-type attack,
which shows that the ES-based two-party QKD protocol presented by Cabello~\cite{Cabello}
is not secure.
In this section, we show that
the security of Cabello's protocol for the ES-based multiparty quantum communication
can also be threatened by the ZLG-type attack.
The eavesdropping strategy is illustrated in Fig.~\ref{fig:fi1},
and can be described as follows:
Initially,
Eve prepares two Bell states, $(e_1, e_2)_{\rm  E}$ and $(f_1, f_2)_{\rm  E}$.
Thus, the initial state is
$(a_1, a_2)_{\rm  A} \otimes (g_1, g_2, g_3)_{\rm  A}
\otimes (e_1, e_2)_{\rm  E} \otimes (f_1, f_2)_{\rm  E} \otimes
(b_1, b_2)_{\rm  B} \otimes (c_1, c_2)_{\rm  C}$.

\begin{itemize}
\item[(a)] Alice performs a Bell measurement
on one qudit of her Bell state and one qudit of her GHZ state
as in the original Cabello's protocol.

Then the state becomes 
$(g_1-k, a_2+l)_{\rm  A}\otimes (a_1+k, g_2-l, g_3-l)_{\rm A}
\otimes (e_1, e_2)_{\rm  E} \otimes (f_1, f_2)_{\rm  E}
\otimes (b_1, b_2)_{\rm  B} \otimes (c_1, c_2)_{\rm  C}$
for some $k$ and $l$.

\item[(b)] Eve intercepts two qudits transmitted from Alice and keeps them intact,
and then sends one qudit of her Bell state to Bob (Carol)
as in Fig.~\ref{fig:fi1} (b).
Then Bob (Carol) makes a Bell measurement
on the received qudit and one qudit of his (her) Bell state
without being concerned of eavesdropping.

Then the state becomes
$(g_1-k, a_2+l)_{\rm  A}\otimes (a_1+k, g_2-l, g_3-l)_{\rm AEE}
\otimes (e_1+\alpha_1, b_2+\alpha_2)_{\rm EB}
\otimes (f_1+\beta_1, c_2 +\beta_2)_{\rm EC}
\otimes (b_1 -\alpha_1, e_2-\alpha_2)_{\rm  B}
\otimes(c_1-\beta_1,e_3-\beta_2)_{\rm C}$
for some $\alpha_i$ and $\beta_i$.
\item[(c)] After their measurements,
Eve intercepts the qudits transmitted from Bob (Carol),
and measures the Bell state
consisting of one of her qudit and the intercepted qudit from Bob (Carol).
Eve performs the suitable local operation on the qudits
which she intercepts from Alice,
according to the measurement result of the GHZ state
as in the following:
If we let $(e_1+\alpha_1, b_2+\alpha_2)_{\rm E}
\otimes (f_1+\beta_1, c_2 +\beta_2)_{\rm E}$
be the measurement result of the two Bell states,
then Eve applies the local operation
\begin{equation}
I\otimes Z^{\alpha_1+\beta_1}X^{b_2+\alpha_2-e_2}\otimes X^{c_2+\beta_2-e_3},
\label{eq:local_operation}
\end{equation}
where $X$ and $Z$ are the unitary operations defined by
\begin{eqnarray}
X: |j\rangle &\mapsto& \left| j+1 \mod{d} \right\rangle \nonumber\\
Z: |j\rangle &\mapsto& \omega_{d}^j \left|j  \mod{d} \right\rangle.
\label{eq:operation_X_Z}
\end{eqnarray}

Using the following relation
\begin{eqnarray}
I\otimes Z^{p}X^{q}\otimes X^{r}(u , v , w) 
=(u+p, v+q, w+r),
\label{eq:operation_property}
\end{eqnarray}
the resulting state can be described by
$(g_1-k, a_2+l)_{\rm  A}\otimes
(a_1+k+\alpha_1+\beta_1, b_2 -e_2+\alpha_2+g_2-l, c_2-e_3+\beta_2+g_3-l)_{\rm AEE}
\otimes (e_1+\alpha_1, b_2+\alpha_2)_{\rm E}
\otimes (f_1+\beta_1, c_2 +\beta_2)_{\rm E}
\otimes (b_1 -\alpha_1, e_2-\alpha_2)_{\rm  B}
\otimes(c_1-\beta_1,e_3-\beta_2)_{\rm C}$,
after applying the local operation in (\ref{eq:local_operation}).
\item[(d)] Eve returns the two qudits to Alice after the above operations.
Then Alice measures the GHZ state,
and publicly announces the result.

Then the final state becomes
\begin{widetext}
\begin{eqnarray}
(g_1-k, a_2+l)_{\rm  A}&\otimes&
(a_1+k+\alpha_1+\beta_1 , b_2-(e_2-\alpha_2)+g_2-l,c_2-(e_3-\beta_2)+g_3-l )_{\rm Pub}\nonumber\\
&\otimes&(e_1+\alpha_1, b_2+\alpha_2)_{\rm E}
\otimes (f_1+\beta_1, c_2 +\beta_2)_{\rm E}\nonumber\\
&\otimes&(b_1 -\alpha_1, e_2-\alpha_2)_{\rm  B}
\otimes(c_1-\beta_1,e_3-\beta_2)_{\rm C}.
\label{eq:ZLG_final_state}
\end{eqnarray}
\end{widetext}
\end{itemize}
Alice, Bob and Carol
now attempt checking 
the eavesdropping
by comparing random subsequences of their measurement results.
However, legal users cannot verify the presence of eavesdropping
since no errors can be introduced from this particular attack
as seen in
Eq.~(\ref{eq:original_final_state}) and
Eq.~(\ref{eq:ZLG_final_state}).
Moreover, Eve can perfectly obtain information on the keys,
which can readily be shown from Eq.~(\ref{eq:ZLG_final_state}).
Therefore, we can prove that Cabello's multiparty quantum communication
protocol is insecure against the ZLG-type attack.


\section{Secure protocol for multiparty quantum communication}\label{sec:Modified_protocol}
In the preceding section,
we have shown that
Cabello's protocol for multiparty quantum communication
is not secure against a particular eavesdropping attack,
the ZLG-type attack.
In this section we suggest a modified protocol,
which is secure
against not only the basic intercept-resend attack but also the ZLG-type attack.
\subsection{Our modified protocol}\label{subsec:modified_protocol}
\begin{figure}[ch]  
\includegraphics[angle=-90,scale=.90,width=.70\textwidth]{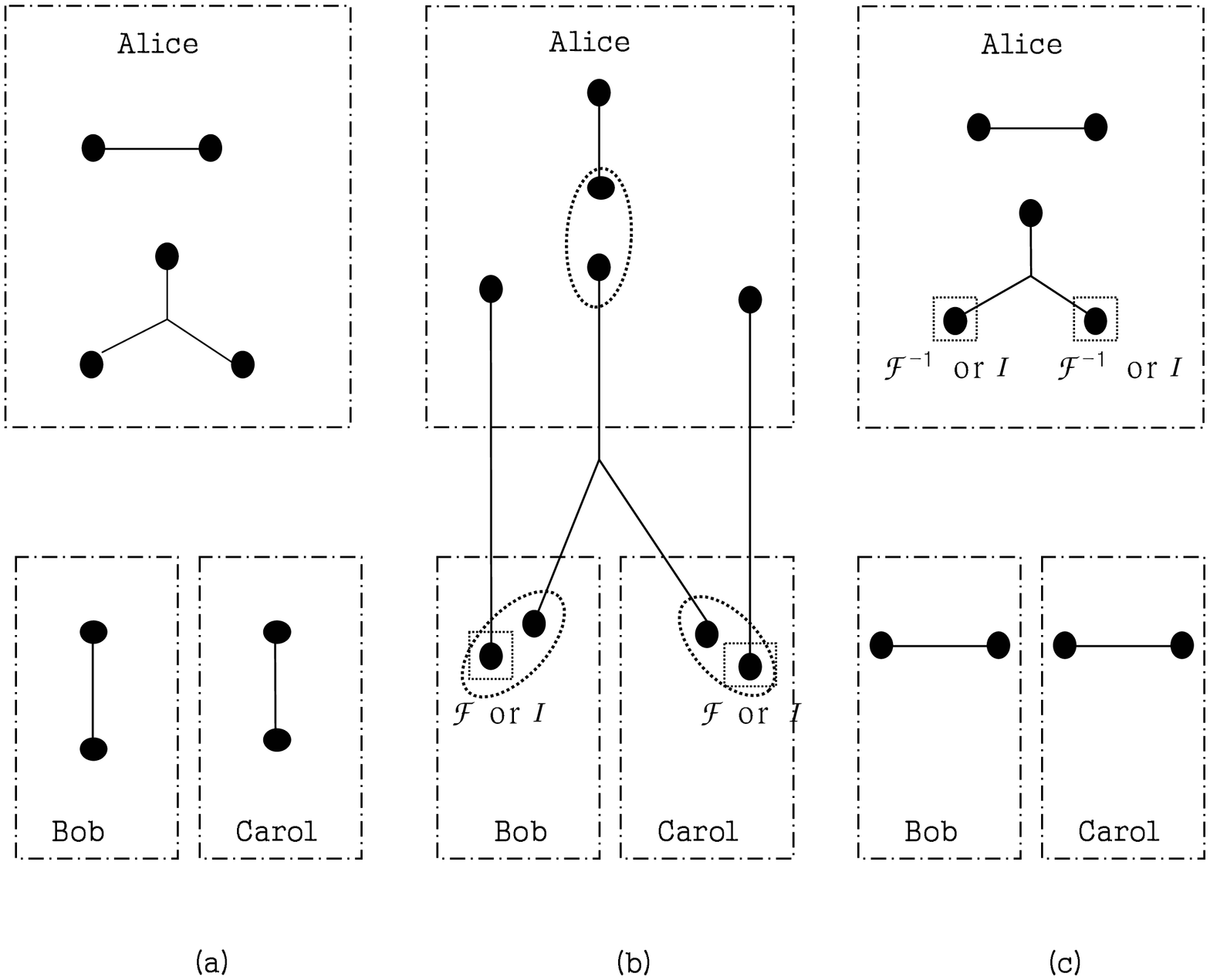}
\caption{\label{fig:fi2} Our modified protocol:
The dotted box represents applying the local operation,
which is randomly the QFT (or the inverse of the QFT) or the identity operation.
In the ZLG-type attack
Eve cannot fabricate the correlation among the keys of legal users
without knowing Bob and Carol's choices of the operation in Step (b).
}
\end{figure}

The basic idea of our modification results from
Cabello's modification for the ES-based two-party QKD~\cite{Cabello2}.
We now remark the following two formulas
\begin{align}
  (u_1, u_2,u_3)_{123} \otimes& [(\mathcal{F}\otimes I)(0 , 0)_{45}] \nonumber\\
   = {\frac{1}{d}}\sum_{k,l=0}^{d-1}\omega_{d}^{-kl}&
   [(I\otimes I\otimes \mathcal{F})(u_1 -k,u_2,l)_{125}]\nonumber\\
   &\otimes(k, u_3 -l)_{43},
 \label{eq:Fourier01}
\end{align}
\begin{align}
  [(I\otimes \mathcal{F}\otimes I)&(u_1, u_2,u_3)_{123}]
  \otimes [(\mathcal{F}\otimes I )(0 , 0)_{45}] \nonumber\\
   = {\frac{1}{d}}\sum_{k,l=0}^{d-1}&\omega_{d}^{-kl}
   [(I\otimes \mathcal{F}\otimes \mathcal{F})(u_1 -k,u_2,l)_{125}]\nonumber\\
   &\otimes(k, u_3 -l)_{43},
 \label{eq:Fourier02}
\end{align}
which can be obtained from the straightforward calculations.
Here, $\mathcal{F}$ is the $d$-dimensional quantum Fourier transform (QFT)
defined by $|k\rangle\mapsto(1/\sqrt{d})\sum_j \omega_d^{jk}|j\rangle$.
The modified protocol is illustrated in Fig.~\ref{fig:fi2}
and is described as follows.
\begin{itemize}
\item[(a)] For convenience, legal users initially prepare the following state:
$(0, 0)_{\rm  A} \otimes (0, 0, 0)_{\rm  A}
\otimes 
(0, 0)_{\rm  B} \otimes (0, 0)_{\rm  C}$.
\item[(b)] Alice sends one qudit of her GHZ state to Bob (Carol).
Bob (Carol) also sends one qudit of his (her) Bell state to Alice,
and then randomly applies either the QFT or the identity operation
to the other qudit of his (her) Bell state.
Alice performs a Bell measurement on her qudit of the GHZ state and one qudit of her Bell state,
and Bob (Carol) also performs a Bell measurement
on the qudit received from Alice and the other qudit of his (her) state.

There exist three cases in this step:
(i) Both Bob and Carol apply the identity operation,
(ii) Only one of them applies the QFT, and
(iii) Both of them apply the QFT.
The case (i) is the same as the case of the original Cabello's protocol.
In the case (ii), if it is Carol who applies the QFT,
and $(-\kappa,\lambda)_{\rm A}$, $(-\mu_1,-\mu_2)_{\rm B}$
and $(-\nu_1,-\nu_2)_{\rm C}$ are the measurement results
of Alice, Bob and Carol respectively,
then Alice's three-particle state becomes
$(I\otimes I\otimes \mathcal{F})
(\kappa+\mu_1+\nu_1, -\lambda+\mu_2 , -\lambda+\nu_2)_{\rm  A}$
by Eq.~(\ref{eq:Fourier01}).
In the case (iii),
if $(-\kappa,\lambda)_{\rm A}$, $(-\mu_1,-\mu_2)_{\rm B}$
and $(-\nu_1,-\nu_2)_{\rm C}$ are the measurement results
of Alice, Bob and Carol respectively,
then Alice's three-particle state becomes
$(I\otimes \mathcal{F}\otimes \mathcal{F})
(\kappa+\mu_1+\nu_1, -\lambda+\mu_2 , -\lambda+\nu_2)_{\rm  A}$
by Eq.~(\ref{eq:Fourier01}) and Eq.~(\ref{eq:Fourier02}).
\item[(c)] Bob (Carol) announces his (her) operation
chosen in the previous step.
Alice performs the inverse of his (her) operation
to the qudit received from him~(her).
Alice now measures the GHZ state
which is automatically made from the legal user's measurements,
and publicly announces the result.
Then the final state becomes
\begin{eqnarray}
&&(-\kappa,\lambda)_{\rm A}\nonumber\\
&&\otimes (\kappa+\mu_1+\nu_1, -\lambda+\mu_2 , -\lambda+\nu_2)_{\rm  Pub}
\nonumber\\
&&\otimes (-\mu_1,-\mu_2)_{\rm B}\otimes (-\nu_1,-\nu_2)_{\rm C}.
\label{eq:modified_final_state}
\end{eqnarray}
\end{itemize}
 For the secure communication, legal users check the eavesdropping
by comparing random subsequences of key strings.

As one can see in Eq.~(\ref{eq:original_final_state}) and Eq.~(\ref{eq:modified_final_state}),
our modified protocol provides legal users
with the same correlation as that in the original Cabello's protocol.
Furthermore,
legal users can detect any eavesdropping
unless Eve knows Bob and Carol's choices of their operations beforehand.
We are now going to discuss the security of our modified protocol in detail.

\subsection{\label{sec:level2}Security check}
As mentioned in the previous section, it is impossible for Eve to get all information
unless she knows Bob and Carol's decisions beforehand.
More precisely,
legal users can detect Eve's presence
with the probability not less than $1-[(d+1)/2d]^2$
for each testing key,
since after applying the QFT to one qudit of a Bell state
one can obtain one of Bell states as the Bell-measurement result
with the probability at most $1/d$.
Thus if legal users check sufficiently large $m$ keys,
with the probability not less than $1-[(d+1)/2d]^{2m}$, Eve can be detected.
Eve can be detected with the probability, not less than
We note that in the $N$-party quantum communication
the probability to detect Eve in the ZLG-type attack,
is at least $1-[(d+1)/2d]^{N-1}$
for each testing key.
Therefore, our modified protocol for multiparty quantum communication
is secure against the ZLG-type attack.

Now, we consider
the basic intercept-resend attacks
on the protocol.
In the three-party quantum communication,
the basic intercept-resend attacks can be divided into two cases,
one-party attack and two-party attack.
While from the one-party attack Eve can acquire the information on the keys only for the QKD,
from the two-party attack Eve can acquire the information on the keys for the QSS as well as the QKD.
In order to investigate the security of the protocol,
we are going to calculate the probability
with which legal users can detect Eve for each case.

\subsubsection{One-party attack}
\begin{figure}[ch]
\includegraphics[angle=-90,scale=0.90,width=.70\textwidth]{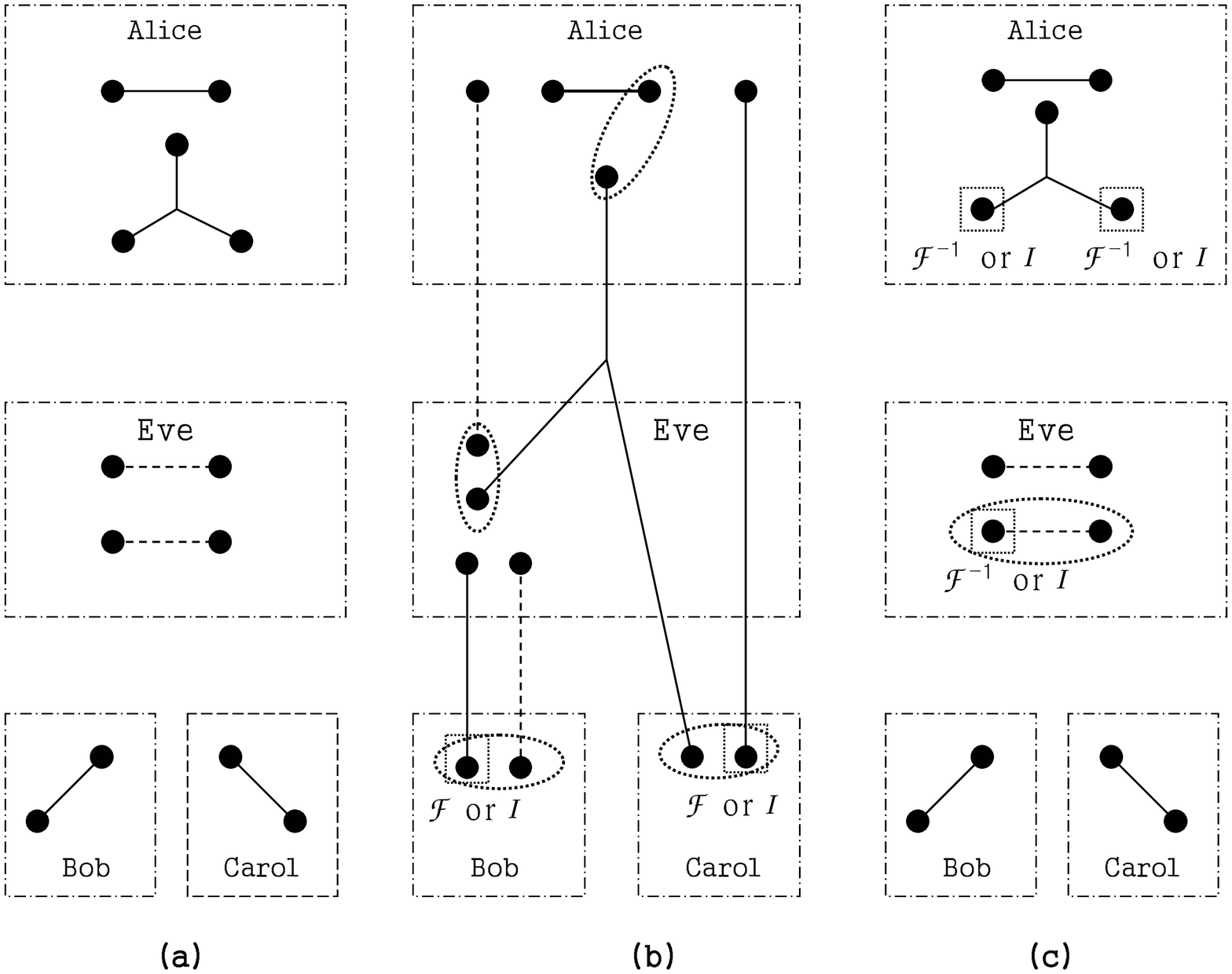}
\caption{\label{fig:fi3} One-party attack on our modified protocol:
 From this attack Eve can acquire the information on the keys
only for the three-party QKD.
}
\end{figure}

The one-party attack on our modified protocol is illustrated in Fig.~\ref{fig:fi3}
and can be described as follows.

\begin{itemize}
\item[(a)] Legal users prepare the same qudits as those in our modified protocol,
and Eve prepares two Bell states
to obtain the information on Bob's key.

\item[(b)] When Alice sends one qudit of her GHZ state to Bob (Carol),
and Bob (Carol) sends Alice one qudit of his (her) Bell state,
Eve intercepts the qudit transmitted from Alice to Bob and
the qudit transmitted from Bob to Alice,
and sends one qudit of her one Bell state to Alice
and one qudit of the other Bell state to Bob.
Bob (Carol) randomly applies either the QFT or the identity operation
to the other qudit of his (her) Bell state.
Alice and Bob (Carol) perform Bell measurements as in the our modified protocol.
While Alice and Bob (Carol) perform Bell measurements,
Eve also performs a Bell measurement on the qudit
which she intercepts from Alice and her qudit of the Bell state
as in Fig.~\ref{fig:fi3} (b).

\item[(c)] After Bob (Carol) announces his (her) operation chosen in the previous step,
Eve performs the inverse of his operation and then measures a Bell measurement
on the state containing the qudit.
\end{itemize}
Finally, legal users check the correlation among their measurement results.
Eve's presence can be detected
since
Eve cannot fabricate the correlation among the legal users' keys
without knowing Bob and Carol's choices of the operations in Step (b).
We now remind that
if one applies the QFT to one qudit of a Bell state (or a GHZ state)
then with the probability at most $1/d$
one can get one of Bell states (or GHZ states) using a Bell-state measurement (or a GHZ-state measurement).
Thus,
we can easily obtain that
the probability of revealing the presence of Eve,
who attempts the one-party attack,
is not less than $1-(d+1)/2d^2$
for each testing key.

In the $N$-party quantum communication,
the probability to detect Eve for the one-party attack
is also at least $1-(d+1)/2d^2$.
Thus our modified protocol is secure against the one-party attack
since the probability for $m$ testing keys is not less than $1-[(d+1)/2d^{2}]^{m}$.
Furthermore, we can obtain from these probabilities
the fact that the higher the dimension of quantum system is, the
more secure the communication becomes.

\subsubsection{Two-party attack}
\begin{figure}[cht]
\includegraphics[angle=-90,scale=.900,width=.70\textwidth]{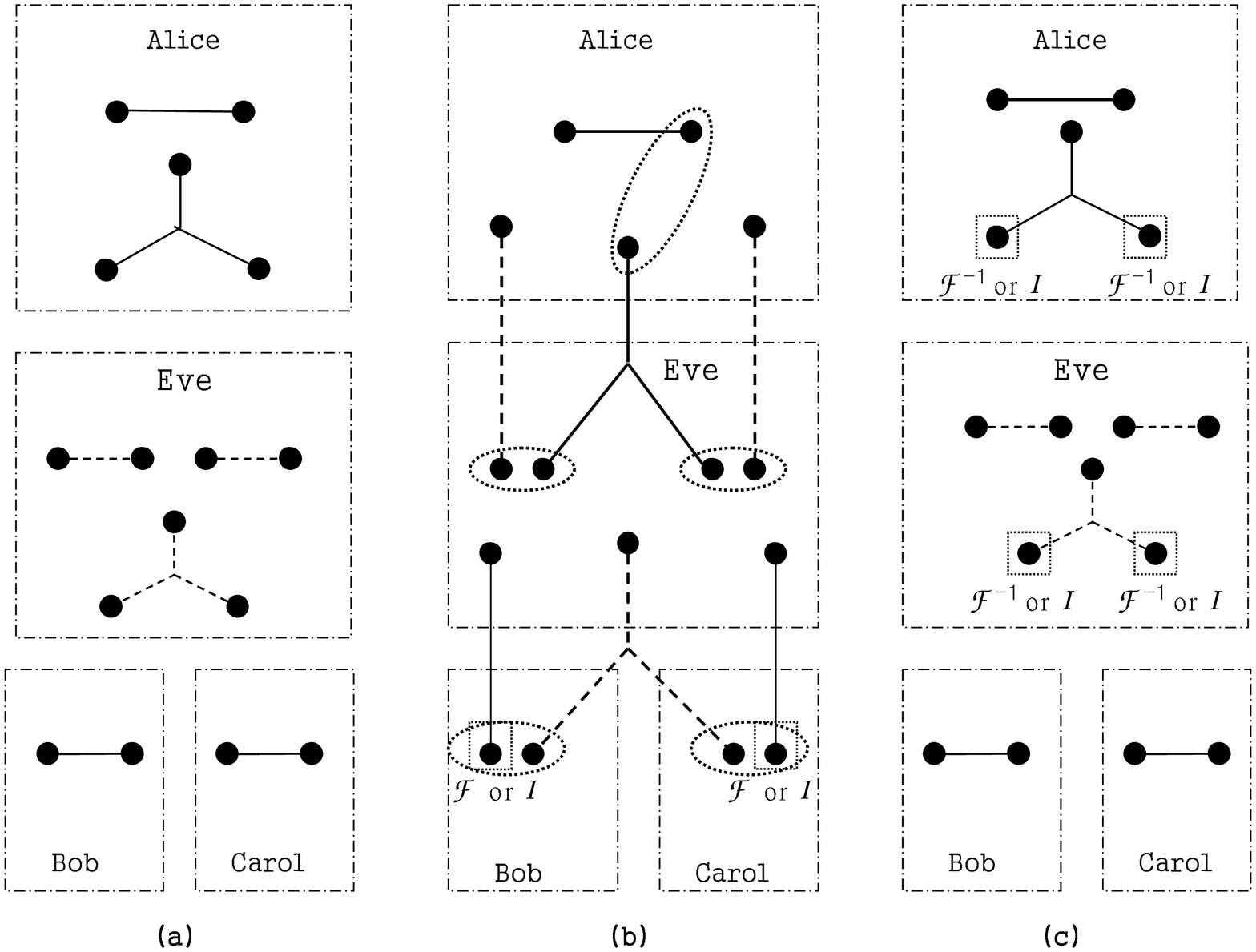}
\caption{\label{fig:fi4} Two-party attack on our modified protocol:
 From this attack Eve can acquire the information on the keys
for the QSS as well as the QKD.}
\end{figure}

The two-party attack on our modified protocol is illustrated in Fig.~\ref{fig:fi4}
and is described as follows.
\begin{itemize}
\item[(a)] The initial states of legal users are the same as those in our modified protocol.
Eve prepares two Bell states and one GHZ state.
\item[(b)] When Alice sends one qudit of her GHZ state to Bob (Carol),
and Bob (Carol) sends Alice one qudit of his (her) Bell state,
Eve intercepts the transmitted qudits, and sends her qudits to legal users instead;
she sends one qudit of each Bell state to Alice
and one qudit of her GHZ state to Bob (Carol).
Bob (Carol) randomly applies either the QFT or the identity operation
to the other qudit of his (her) Bell state.
Alice, Bob and Carol perform Bell measurements as in the our modified protocol.
While they perform Bell measurements,
Eve performs two Bell measurements
on her own qudits and the qudits
which she intercepts from Alice
as in Fig.~\ref{fig:fi4} (b).
\item[(c)]After Bob (Carol) announces his (her) operation chosen in the previous step,
Eve performs the inverse of his (her) operation
to the qudit received from him~(her),
and then measures the GHZ state containing the qudits.
\end{itemize}
In this eavesdropping strategy,
Eve cannot forge the exact correlation on the keys
for the three-party quantum communication
without knowing Bob's (Carol's) decision to choose his (her) own operation.
Thus, 
legal users can detect Eve's intervention
by publicly comparing a sufficiently large random subset of their sequences of key strings.
We can straightforwardly show that the probability with which Eve can be detected
in the three-party quantum communication,
is at least
\begin{equation}
1- \frac{2d^2+d+1}{4d^3}
\label{eq:3-party_2p_Prob}
\end{equation}
 for each testing key.
 Thus our modified protocol is secure against the two-party attack
since the probability for $m$ testing keys is not less than $1-[ {(2d^2+d+1)}/{4d^3}]^{m}$.

In order that Eve acquire the keys for the QSS in the multiparty quantum communication,
she should access the qudits transmitted to all legal users.
Thus, employing Eq.~(\ref{eq:3-party_2p_Prob}),
one can obtain that the probability of detecting Eve's presence
in the $N$-party quantum communication is not less than
\begin{equation}
1-\frac{d(d+1)^{N-1}-d^N+1}{2^{N-1}d^{N}},
\label{eq:N-party_2p_Prob}
\end{equation}
for each testing key.

It follows from the above probability in Eq.~(\ref{eq:N-party_2p_Prob})
that as the dimension of the system or the number of parties increases,
the multiparty quantum communication including the multiparty QKD and QSS
is more secure against the two-party attack.

\section{Conclusions}\label{sec:Conclusions}
In this paper, we presented the ZLG-type attack applied to Cabello's protocol
for the multiparty quantum communication
consisting of the multiparty QKD and the QSS of classical information,
and showed that the protocol is not secure against the attack.
Modifying the protocol,
we also showed that there exists a multiparty quantum communication protocol
which is secure against the ZLG-type attack
as well as the basic intercept-resend attack.

Cabello's ES based protocol \cite{Cabello1} simultaneously makes
the multi-party QKD and the QSS of classical information possible.
Furthermore, it has a merit that no transmitted quantum data are rejected
as mentioned in \cite{Cabello1}.
Our work shows that we can obtain the secure protocol with those advantages
by exploiting an appropriate modification process.
\begin{acknowledgments}
The authors acknowledge the KIAS Quantum Information Group for
useful discussions.
S.L. is supported by a KIAS Research Fund (No.~02-0140-001),
J.K. by a Korea Research Foundation Grant (KRF-2002-070-C00029),
and S.D.O. by a Korea Science and Engineering Foundation Grant (R06-2002-007-01003-0).
\end{acknowledgments}


\end{document}